\documentclass{aa}  
\usepackage{graphicx}
\usepackage{txfonts}
\usepackage{lipsum}
\usepackage{subcaption}  
\usepackage{lscape} 
\usepackage{hyperref}
\usepackage{natbib}
\usepackage{CJK}
\usepackage[utf8]{inputenc}

\newcommand   \K     {\rm K_{S}}
\newcommand   \Wi    {\rm W1}
\newcommand   \Aadd  {\mathit{A}_{\rm W1}^{\rm ice}}

\authorrunning{Cao et al.}
\begin{document}
\begin{CJK*}{UTF8}{gbsn}
\defcitealias{Cao24}{Paper\ I}
\defcitealias{stefan17_PNICER}{PNICER}

\title{The First Glimpse of Water Ice Absorption Map in the Milky Way}

\author{Zhetai Cao (曹哲泰) \inst{1,2,3}
\and Biwei Jiang (姜碧沩) \inst{1,3}
\and Stefan Meingast\inst{2}}

\institute{School of Physics and Astronomy, Beijing Normal University, Beijing 100875, China
\and University of Vienna, Department of Astrophysics, T\"urkenschanzstra{\ss}e 17, 1180 Vienna, Austria
\and Institute for Frontiers in Astronomy and Astrophysics, Beijing Normal University, Beijing 102206, China}

\abstract
{Interstellar ice plays a key role in the thermal evolution of the interstellar medium and in astrochemical pathways, yet its large-scale distribution remains poorly constrained. We use ALLWISE and 2MASS photometry to estimate water ice absorption in the $\Wi$/WISE band by correcting the observed colors for reddening and intrinsic stellar colors. This allows us to construct a first Milky Way water ice map. By varying input parameters, we test the stability of the method and identify the extinction law as the dominant source of uncertainty. Using synthetic photometry, we also quantify how different physical and observational parameters influence the $\Wi$ band water ice absorption. The strong correlation between the measurement from photometric method and spectroscopic water ice abundance confirms that the $\Wi$ band signature originates from the 3 $\mu$m ice feature. We present the relationship between ice absorption in $\Wi$ band and water ice optical depth from theory and observations. Finally, we provide a preliminary Milky Way-scale map of water ice distribution.
}


\maketitle
\section{Introduction}
Interstellar ice is an important probe of evolution history and chemistry of the interstellar media (ISM), as it participates in many astrophysical processes. During the past decades, many studies have been conducted to study the composition and abundance of the ice species, and most identifications of ice are made by matching the absorption in infrared (IR) observation spectrum with laboratory derived profile. Nowadays, various molecule species have been identified, ranging from simple molecules, such as water, $\rm CO_{2}$ and CO, to complex organic molecules, such as Polycyclic Aromatic Hydrocarbons (PAHs) \citep{Boogert15_ARAA, McClure23_IcyAge, Ewine09_ARA&A_COM, Gibb04_IRLegacy, berg11_SpitzerLegacy}. Ice is widely found in various environments, from dense clouds \citep{Boogert11_Spec_DenseCore, Whittet13_Spec_L183} to areas surrounding YSOs \citep{Rocha24_JOYs, Boogert08_c2dYSO} and even the Galactic Center region \citep{Ginsburg23_CO, Chiar00_IceGC}. For these various ices in the universe, one of the most obvious absorptions in the spectrum is from the O-H stretching modes in water ice molecule, which is one of the most prominent absorption features in IR spectrum around 3 $\mu$m \citep{Boogert15_ARAA}.

There are some studies focusing on the study of 3 $\mu$m water ice about the abundance, the correlation with other molecules and the interaction with surrounding environment \citep{Madden22_Spec_PerSer, Whittet13_Spec_L183, Boogert13_Spec_Lupus, Boogert11_Spec_DenseCore}. Most of the studies rely on the spectrum data. However, the observations of IR spectrum near 3 $\mu m$ are relatively rare due to the strong IR absorption from The Earth's atmosphere and the limited technology in IR telescope. In addition, spectroscopic observation is hard to carry out in a large areas of sky, making it challenging to study a large-scale distribution of ice. However, some filters may enable us to detect ice by photometric rather than spectroscopic methods. \citet[hereafter \citetalias{Cao24}]{Cao24} reported an offset in $\Wi$/WISE band when studying the extinction law in nearby molecular clouds, which was interpreted as the effect of water ice absorption near 3 $\mu$m. \citet{Ginsburg23_CO} uses narrow filters of JWST to trace the absorption from CO ice in central molecular zone near the Galactic Center. Typically, stars should generally align parallel to the extinction vector in the Color-Color Diagram (CCD) if dust extinction is the only effect. However, ice absorption introduces additional extinction that shifts stars away from this direction. This shift is clearly visible in Figure 4 in \citet{Ginsburg23_CO}, which demonstrates the existence of CO ice in the region. The recent paper from \citet{Stefan25_Ice_arXiv} provides a detailed discussion on the method of using the $\Wi$/WISE or IRAC1/Spitzer (hereafter, I1/Spitzer) band to detect water ice absorption features in the IR. These results imply that photometry is able to detect the molecule absorption efficiently and easily and provide us a possibility to study the large-scale ice distribution. 

Here, we use W1/WISE filter to trace water ice. By removing the effect of extinction and intrinsic color, the additional absorption is calculated and used to construct the 2D distribution of water ice in Milky Way. In Section \ref{sec:Data}, a preliminary data quality control is introduced. In Section \ref{sec:Calculation}, the procedure for calculating water ice absorption from the observed color, extinction, and intrinsic color is described in detail. In Section \ref{sec:Influence}, we analyze the influence of the extinction law on the photometric ice absorption calculation. In Section \ref{sec:SynPhot}, using synthetic photometry, we theoretically investigate the influence from absorption profile and extinction. In Section \ref{sec:Verification}, photometric ice absorption results are compared with spectroscopic observations results. The relationship between ice absorption in magnitude and the water ice abundance are presented. In Section \ref{sec:Result} we provide the global and detailed maps of water ice distribution. In Section \ref{sec:Discussion}, the expectation of photometric ice method is discussed. Section \ref{sec:Summary} is the summary of the paper.

\section{Data}\label{sec:Data}
To calculate the additional absorption in W1 band, we use all the data from the Wide-field Infrared Survey Explorer (WISE) survey, which is a full sky survey covering four bands, W1, W2, W3, W4 with effective wavelength at 3.35, 4.60, 11.56 and 22.09 $\mu$m, respectively \citep{WISE}. Here, we use W1 and W2 data from ALLWISE catalog. 2MASS project is a whole-sky survey with J, H, $\K$ bands in near-IR with the effective wavelength at about 1.24, 1.66, and 2.16 $\mu$m respectively \citep{2MASS}. Since ALLWISE data has included 2MASS data, no additional cross-match is required. About 700 million stars are included in the dataset.

Furthermore, we applied the selection criteria on the dataset to ensure the quality:
\begin{itemize}
    \item photometry errors of the three 2MASS bands are less than 0.05 mag.
    \item photometry errors of the two ALLWISE bands are less than 0.1 mag.
    \item $ph\_qual = A$ in W1 and W2, which ensures the source is detected in these two bands with a flux signal-to-noise ratio greater than 10.
    \item $ext\_flg = 0$ in W1 and W2, which ensures the source shape is consistent with a point-source and the source is not associated with or superimposed on a previously known extended object in 2MASS Extended Source Catalog.
    \item $w1flg = 0$ and $w2flg = 0$, which ensures pixels within the measurement aperture are not affected by nearby objects, saturated pixels, or other unusable data.
    \item $n\_2mass = 1$, which ensures that only one 2MASS source is found within a 3$\arcsec$ radius of the WISE source position.
    \item $var\_flg \le 5$ in W1 and W2 bands, flag less than 5 are most likely not variables as recommended \citep{WISE}.
\end{itemize}
After the data quality selection, approximately 20 million stars remain in the sample.

\section{Calculation of Water Ice Absorption in $\Wi$ Band}\label{sec:Calculation}
In brief, the observed color of a star is determined by its intrinsic color and dust extinction. In this work, we account for the effect of water ice absorption. The corresponding equation is given as follows:
\begin{equation}\label{eq:obs_expression}
(\K-\Wi)=(\K-\Wi)_{0}+\it{E}_{\K,\Wi}^{\rm{dust}}+\it{E}_{\K,\Wi}^{\rm{ice}}
\end{equation}
($\K-\Wi$) and $(\K-\Wi)_{0}$ are the observed and the intrinsic color of the star, respectively. $E_{\K,\Wi}^{\rm{dust}}$ and $E_{\K,\Wi}^{\rm{ice}}$ are the reddening from interstellar dust and water ice absorption, respectively. 

\begin{figure}[h!]
    \centering
    \includegraphics[width=\hsize]{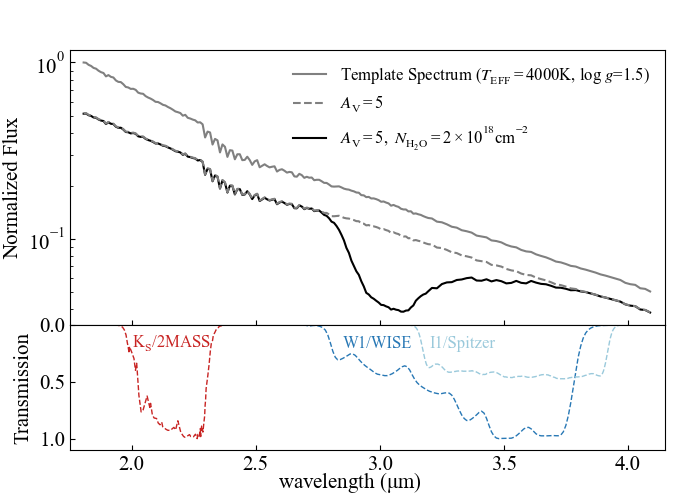}
    \caption{The model-predicted water ice absorbed spectrum of a star with an assumed extinction of $A_{V}=5$ mag from WD01 dust model \citep{WD01} and $\it{N}_{\rm{H_{2}O}}\rm{=20\times10^{17}cm^{-2}}$. The different lines mark the scaled spectrum from ATLAS9 \citep{ATLAS9}, the spectrum under dust extinction and the spectrum under dust extinction mixed with water ice absorption. The transmission curves of $\K$/2MASS, W1/WISE and I1/Spitzer are shown for comparison with spectrum.}
    \label{fig:model}
\end{figure}

\begin{figure}[h!]
    \centering
    \includegraphics[width=\hsize]{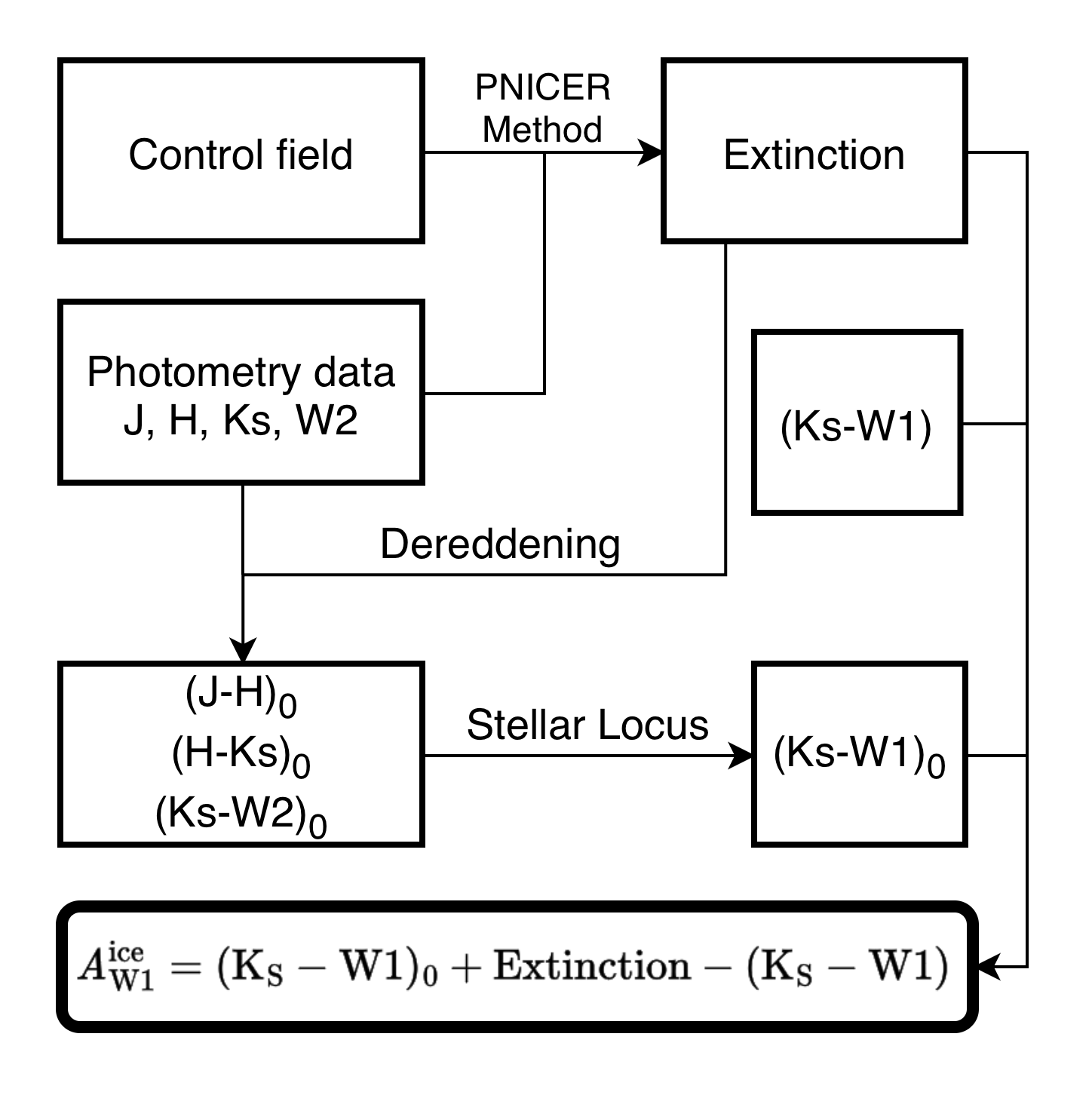}
    \caption{Flowchart of the photometric water ice calculation.}
    \label{fig:flowchart}
\end{figure}

In Figure \ref{fig:model}, the transmission of several IR bands and the stellar spectrum with water ice absorption are shown, since the 3 $\mu$m water ice profile does not overlap with $\K$ band transmission, it can be inferred that:
\begin{equation}\label{eq:ice_part}
E_{\K,\Wi}^{\rm{ice}} = A_{\K}^{\rm{ice}}-A_{\Wi}^{\rm{ice}} = - A_{\Wi}^{\rm{ice}}
\end{equation}
where $A_{\K}^{\rm{ice}}$ and $A_{\Wi}^{\rm{ice}}$ refer to the extinction caused by ice in these two bands. Then, Equation \ref{eq:obs_expression} becomes:
\begin{equation}\label{eq:ice_calculation}
\Aadd=(\K-\Wi)_{0}+\it{E}_{\K,\Wi}^{\rm{dust}}-(\K-\Wi)
\end{equation}
$\Aadd$ represents the extinction from water ice absorption in W1 band. $(\K-\Wi)$ is the observed colors of each star (in Section \ref{subsec:ObservedColor}). $E_{\K,\Wi}^{\rm{dust}}$ (hereafter $E_{\K,\Wi}$) is the extinction calculated from PNICER method (in Section \ref{subsec:PNICER}). $(\K-\Wi)_{0}$ is the intrinsic color which is calculated from the CCD of `extinction-free' stars (in Section \ref{subsec:IntinsicColor}). The calculation procedure is illustrated in the flowchart shown in Figure \ref{fig:flowchart}. Briefly, we use multi band photometry together with extinction law to derive the extinction $E_{\K,\Wi}$. The stars are then dereddened in $\rm{J-H}$, $\rm{H-\K}$, and $\rm{\K - W2}$ using the derived extinction. The intrinsic $\K - \Wi$ color is subsequently interpolated using the stellar locus method. Finally, $\Aadd$ is derived with Equation \ref{eq:ice_calculation}. 

\subsection{$(\K-\Wi)$: Observed Color Index and Error}\label{subsec:ObservedColor}
The observed color index and its error are given by:
\begin{equation}
(\K-\Wi) = \it{m}_{\K}-\it{m}_{\Wi}
\label{eq:obs_color}
\end{equation}
\begin{equation}    
\sigma_{(\K-\Wi)}=\sqrt{\sigma_{m_{\K}}^{2}+\sigma^{2}_{m_{\Wi}}}
\label{eq:obs_color_err}
\end{equation}
$m_{\K}$, $\sigma_{m_{\K}}$, $m_{\Wi}$ and $\sigma_{m_{\Wi}}$ are the photometry magnitude and its photometry error in $\K$ and $\Wi$ band.

\subsection{$E_{\K,\Wi}$: Extinction and Error from PNICER}\label{subsec:PNICER}
To acquire the extinction of each star, we adopt the PNICER method from \citet[hereafter \citetalias{stefan17_PNICER}]{stefan17_PNICER}, which is an unsupervised machine-learning technique for calculating extinctions. An `extinction-free' control area is required to provide the distribution of intrinsic colors of the stars, then, Gaussian Mixture Models will fit along the extinction vector to derive probability densities that describe the extinction for each star in the CCD. This method does not require any prior knowledge, only observed colors and a given extinction law are sufficient. It has been applied in several works to determine the extinction of stars to calculate extinction or construct extinction maps \citep{Zhang23_PNICERmap, stefan18_Orion}. \texttt{PNICER} Python package (\url{https://github.com/smeingast/PNICER}) is used to calculate extinction and its error. 

\begin{figure*}[h!]
    \centering
    \includegraphics[width=\hsize]{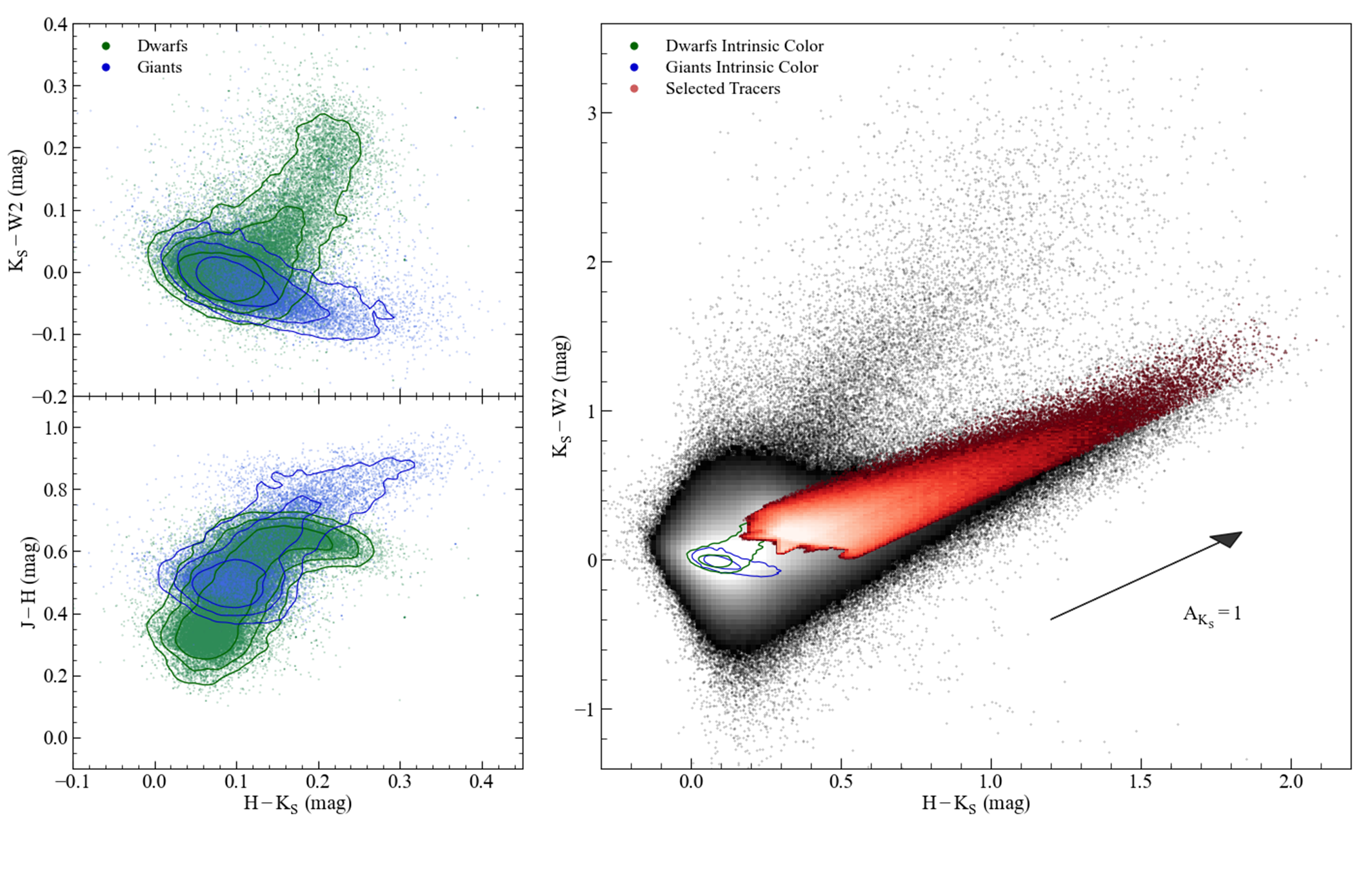}
    \caption{The intrinsic color-color diagram of H-$\K$/$\K$-$\rm{W2}$ (upper left) and H-$\K$/J-H (lower left) and the CCD from ALLWISE dataset. The left two panels show the CCD of `extinction-free' control samples from \citetalias{Cao24}, where the blue and green points represent the intrinsic colors of giants and dwarfs respectively, and contour maps represent the density distribution. The right panel displays the density map of the 20 million ALLWISE sources (shown in a binary color scale). The arrows indicate the direction of the extinction vector for $A_{\K}=1$. The blue and green contours are the same as those in the left panels. The final sample of stars used to trace the ice distribution is marked in red.}
    \label{fig:IC_CCD}
\end{figure*}

In order to derive extinction, the input photometry data come from the selected data in Section \ref{sec:Data}. In addition, to avoid the influence from absorption in W1 band, only J, H, $\K$ and W2 bands, three colors, $\rm{J-H}$, $\rm{H-\K}$ and $\rm{\K-W2}$, are used to estimate extinction. The extinction law used here is $(A_{\rm{J}}/A_{\K}, A_{\rm{H}}/A_{\K}, A_{\rm{W2}}/A_{\K})=(2.880,1.637,0.414)$ \citepalias{Cao24}. The `extinction-free' samples are taken from \citetalias{Cao24}. Briefly, \citetalias{Cao24} aims to study the extinction law in different ISM environment. A large number of stars with precise extinction in IR band are derived by combining the stellar parameters from the near-IR spectroscopic survey APOGEE \citep{APGDR17} with photometric data from 2MASS, WISE, and Spitzer. In this paper, we use the average extinction law and the derived extinction data. Here, we define giants stars that satisfy $E_{\mathrm{J},\K} < 0.02$ in \citetalias{Cao24} as `extinction-free' control samples. The reasons why only giants are used are explained in Appendix \ref{App:tthSet}. Briefly, dwarfs cannot effectively trace high extinction regions and low extinction dwarfs are likely to be misidentified as high extinction giants, which results in a wrong estimation of extinction and $\Aadd$. The CCD of the input control sample is shown in the left panel in Figure \ref{fig:IC_CCD}, the blue points and contour are the input data and its density distribution in the CCD. The output parameters are $A_{\K}$ and $\sigma_{A_{\K}}$. Using the given extinction law, $A_{\K}$ can be converted into the extinction in any other band or color. For example, $E(\K-\Wi)/A_{\K}=1-A_{\Wi}/A_{\K}=0.419$ \citepalias{Cao24}.

\begin{figure}
    \centering
    \includegraphics[width=\hsize]{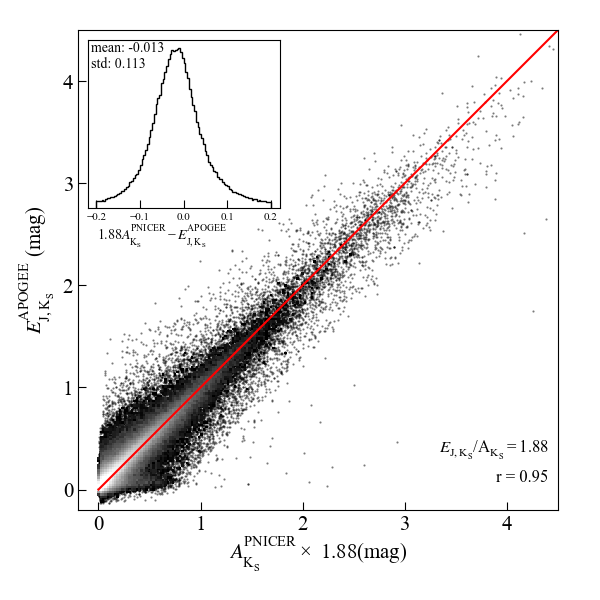}
    \caption{The comparison between extinction derived from PNICER method and that obtained from spectroscopically determined stellar parameters \citetalias{Cao24}. The extinction law to convert $E_{J,\K}$ to $A_{\K}$ is taken from \citetalias{Cao24}. The inset shows the distribution of the difference between the results from two methods. The red line is the identity line $y=x$ and correlation coefficient ($r=0.95$) is marked in the right bottom corner.}
    \label{fig:APOGEECompare}
\end{figure}

We also compare the extinction from PNICER method with the extinction calculated in \citetalias{Cao24} from blue-edge method to verify the reliability of PNICER results. The comparison samples satisfy the criterion $A_{\K}>3\sigma_{A_{\K}}$, leaving 243k stars, as shown in Figure \ref{fig:APOGEECompare}. The results derived from two different methods are highly consistent, with the mean and standard deviation of their differences being $-0.013$ and 0.113, respectively.

Extinction is the most critical parameter in the subsequent calculations. Some selections are applied to improve the data quality. The extinction measurement is constrained by the quality of photometry data and the limitation of the method itself. The errors from \texttt{PNICER} package are mainly $\sigma_{A_{\K}}<0.01$, reflecting only the effect of the intrinsic dispersion of the stellar locus derived from the control sample. The photometric uncertainties are approximately 0.02 to 0.03 mag in each of the three 2MASS bands and about 0.01 mag in the two WISE bands. As a result, the following selection $A_{\K}>3\sigma_{A_{\K}}$ is applied. In addition, to minimize the misidentification between high extinction giants and low extinction dwarfs, we applied an extinction threshold $A_{\K}>0.3$ to mitigate the influence of dwarfs. Details are in Appendix \ref{App:tthSet}. Finally, the following criterion is adopted,
\begin{equation}\label{eq:AkSelection}
A_{\K}>max(0.3,\ 3\sigma_{A_{\K}}),(A_{\K}/A_{\rm V}\approx0.08-0.11)
\end{equation}

\subsection{$(\K,\Wi)_{0}$: Intrinsic Color and Error}\label{subsec:IntinsicColor}
To estimate the intrinsic colors and their uncertainties for each star, the stellar locus method is applied. This method makes use of the empirical relations among stellar colors in color$-$color space to predict one color from others \citep{Map_Chen14, Locus_Xu22}. In this work, the locus is defined by an `extinction-free' control sample, and the intrinsic color, $(\K-\Wi)_{0}$, is inferred from the following three intrinsic colors, $\mathrm{(J-H)_{0}}$, $\mathrm{(H-\K)_{0}}$ and $\mathrm{(\K-W2)_{0}}$. The three intrinsic colors are derived from photometry with dereddening. Specifically, for a given tracer star and its observational color index vector in multi color $\vec{S}$, we derive its three intrinsic colors $\vec{S}_{0}$, $\mathrm{(J-H)_{0}}$, $\mathrm{(H-\K)_{0}}$ and $\mathrm{(\K-W2)_{0}}$ by dereddening. Then, with \texttt{NearestNeighbors} algorithm from \texttt{scikit-learn} \citep{scikit-learn}, the 20 stars closest to the intrinsic colors vector $\vec{S}_{0}$ in the color-color space are selected. These 20 neighboring stars in control sample are taken to represent the intrinsic color of the star. The mean and standard deviation of the $\K-\Wi$ of the 20 selected control sample stars are adopted as the intrinsic color and its uncertainty of the estimation of intrinsic color in $\K-\Wi$.

However, in practice, differences between the intrinsic colors of dwarfs and giants, errors in the extinction estimation, and other factors may cause the derived intrinsic colors to deviate significantly from the stellar locus defined by the control sample. Since a higher quality sample is more preferred for map construction, tracer stars whose derived intrinsic colors significantly deviate from the stellar locus are excluded. To quantify this, we introduce another parameter, $\overline{D}_{\mathrm{neighbors}}$, defined as the average distance in color color space between $\vec{S}_{0}$ and its 20 nearest neighboring stars. A large $\overline{D}_{\mathrm{neighbors}}$ indicates that the star's intrinsic color may deviate from the locus, implying a possible mistake in either the intrinsic color or extinction estimation. Stars should follow the criterion:
\begin{equation}\label{eq:IntrinsicSelection}
\overline{D}_{\rm{neighbors}}<0.05
\end{equation}

\subsection{Final Ice Tracers Samples}\label{subsec:criterion}
In the right panel of Figure \ref{fig:IC_CCD}, the red color shows the result distribution after the criteria above (extinction quality selection from Equation \ref{eq:AkSelection} and $\overline{D}_{\mathrm{neighbors}}$ selection from Equation \ref{eq:IntrinsicSelection}). The maximum extinction is limited to around $A_{\K}\approx3$. About 618k tracers are kept to establish the ice distribution map. 

\section{Influence of Extinction Law on $\Aadd$ Calculation}\label{sec:Influence}

\begin{figure}
    \centering
    \includegraphics[width=\hsize]{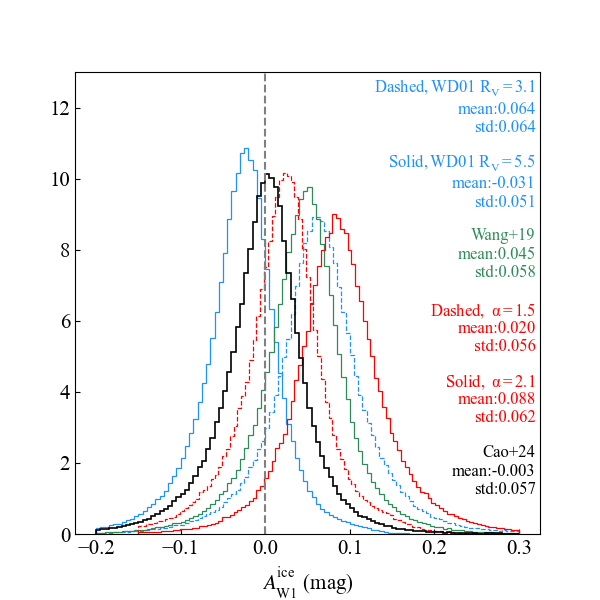}
    \caption{Distribution of $\Aadd$ by keeping the control sample and photometric data fixed while varying only the extinction law. Colors and line styles represent different extinction laws, including the WD01 grain model \citep{WD01}, Wang+19 \citep{Wang19_Law}, Cao+24 \citepalias{Cao24}, and the classical near-infrared power-law \citep{Wang_IRLawJWST}. The mean and standard deviation of each distribution are shown on the right side.}
    \label{fig:ResDistribution}
\end{figure}

We tested several other extinction curves for comparison, including WD01 \citep{WD01}, Wang+19 \citep{Wang19_Law}, and the classical near IR power law. WD01 model establishes a dust grain model with different values of $R_{\rm{V}}$, while Wang+19 derived an extinction law for the Milky Way using multi band photometry. In addition, the classical NIR power law is expressed as $A_{\lambda}/A_{\K}\propto \lambda^{-\alpha}$. The extinction laws are discussed latter and shown in upper panel of Figure \ref{fig:Ext_and_Law}. Using the methods and criteria described in Section \ref{sec:Calculation}, $\Aadd$ under different extinction laws are derived and shown in Figure \ref{fig:ResDistribution}. The histogram of $\Aadd$ values are presented with the mean and standard deviation of each distribution indicated on the right hand side.

\subsection{Choice of the Extinction Law}
The choice of extinction law is important since it has the largest impact on the $\Aadd$. The distribution of derived $\Aadd$ under various extinction laws are shown in Figure \ref{fig:ResDistribution}. Systematic offsets are evident, which originating from differences in the extinction curves. Extinction curves significantly affect the derived extinction and consequently propagate into $\Aadd$. The dispersion of the distribution reflects the measurement uncertainties, which primarily arise from photometric errors and intrinsic color estimation errors. The extinction law, Cao+24, with a smaller offset is preferred in this paper. 

We further test different extinction laws and compare the $\Aadd$ with spectroscopic observation results described in Section \ref{sec:Verification}. Specifically, we compared the $\Aadd$ derived from photometry with the spectroscopic $\tau_{3.0}$ under different extinction curves. A simple linear fit is performed, and the Pearson correlation coefficient $r$ is calculated. For the extinction laws of Cao+24, Wang+19, WD01 ($R_{\rm{V}}=3.1$ and $R_{\rm{V}}=5.5$), and the power laws with $\alpha=1.5$ and $\alpha=2.1$, the $r$ are 0.88, 0.89, 0.82, 0.77, 0.86, and 0.79, respectively. The corresponding intercepts are 0.00, --0.05, --0.06, 0.09, --0.03, and --0.03. These results indicate that the extinction law primarily affects the slope and intercept of the correlation, while the correlation coefficients remain nearly unchanged. This suggests that the extinction law has a measurable influence on $\Aadd$. It does not affect the detection of water ice or relative absorption depth since it affect the zero point of $\Aadd$. The spectroscopic results also support the use of the extinction curve from  Cao+24, as it yields the smallest intercept.

\subsection{Uncertainty Estimation from Extinction Law}
The IR extinction law remains under debate. The extinction law can vary with both the line of sight and the extinction strength \citep{McClure09, Zasowski09, stefan18_Orion, Oph_law_Li23}. A comprehensive IR extinction law map is not yet available since most extinction law maps are based on optical data and can not be applied to dense regions in IR. As a result, it is difficult to estimate the uncertainty introduced by extinction law for individual stars. However, the uncertainties from extinction laws on $\Aadd$ can still be estimated by comparing the result $\Aadd$ from two significantly different extinction laws. This provides an approximate assessment of how variations in the extinction law under different ISM environments may affect the derived $\Aadd$, even if there is no knowledge about the distribution of ISM properties in IR.

Several extinction curves with substantial differences are adopted in this section, including WD01, classical power law near IR extinction law, as well as the average Milky Way extinction law derived from red clump stars. As is shown in Figure 4. In the WD01 models, the difference between $R_V = 5.5$ and $R_V = 3.1$ is 0.095, which represent a flat and steep extinction curve, respectively. In the power law models, the difference between $\alpha = 2.1$ and $\alpha = 1.5$ is 0.068. Conservatively, the uncertainty from extinction law is about 0.03 to 0.05, based on the variations among the WD01 and power-law extinction models. Given that regions with ice absorption are typically dense ISM regions, which generally favor larger $R_V$ values (flatter extinction curves), the actual uncertainty is likely to be a bit smaller.

\section{Model Derived Synthetic $\Wi$ Band Water Ice Absorption}\label{sec:SynPhot}
In this section, we use ATLAS9 stellar model spectra \citep{ATLAS9} combined with different extinction conditions (including various extinction laws and extinction levels) and different water ice absorption models to derive the relationship between the optical depth, $\tau_{3.0}$, of water ice absorption at 3 $\mu$m, and $\Aadd$. Specifically, we use the \texttt{synphot} Python package \citep{synphot} to compute synthetic photometry for the $\K$ and $\Wi$ bands of the original spectra, the spectra with different levels of extinction or different extinction laws, and the spectra with both extinction and water ice absorptions. By comparing the results among different cases, we quantify the relative contribution of each component to the overall absorption. In this section, $\Aadd=m_{\Wi}^{\mathrm{dust+ice}}-m_{\Wi}^{\mathrm{dust}}$, in which `$m$' represents the magnitude.

By using a controlled-variable approach, we study which factors have the greatest influence on $\Aadd$. The intrinsic spectrum is fixed in this part. We try different sets of stellar parameter, but the resulting differences in the synthetic $\Aadd$ are very small. Another reason for the choice is that, since most of our tracers are red clumps, the adopted parameter set should reflect the stellar parameter of the red clumps. That is the reason why we choose stellar parameters with $\rm{T}_{eff}=4250$K, $\rm log\ {\it g}=2.5$, [M/H]=0.0.

\subsection{3.0 $\mu$m Ice Profile}\label{subsec:profile}
The main absorption feature of water ice occurs around 3.0 $\mu$m and is primarily due to the O$-$H stretching modes of water molecules, and the profile shows little variation in most situations. In addition to this dominant feature, several weaker structures appear on both the blue and red sides near 3.0 $\mu$m. The blue-side wing near 2.9 $\mu$m is less commonly observed \citep{Boogert15_ARAA}. On the red side, however, some components are used to explain the extended red wings. One corresponds to molecular absorption, C$-$H stretching mode in methanol $\rm{CH_{3}OH}$ at 3.53 $\mu$m and $\rm{NH_{3}}$ related species around 3.47 $\mu$m. The other is a broad, extended absorption which is thought to originate from Mie scattering by large grains \citep{Boogert15_ARAA}. In observations, the extended red wing exhibits various types. \citet{RedWings_Noble13} classified the wings into three types according to the strength. Tight correlations have been reported between the strength of the red wing and the water ice column density. Although this correlation may also be coincidental due to the small number of sources in their work, it still provides some evidence that the carriers of the extended wing may be related to water ice \citep{RedWings_Thi06}. This wing is commonly observed in the spectra of background stars with water ice absorption, and its profile shows little variation with extinction \citep{Madden22_Spec_PerSer, Boogert11_Spec_DenseCore}.

\begin{figure}
    \centering
    \includegraphics[width=\hsize]{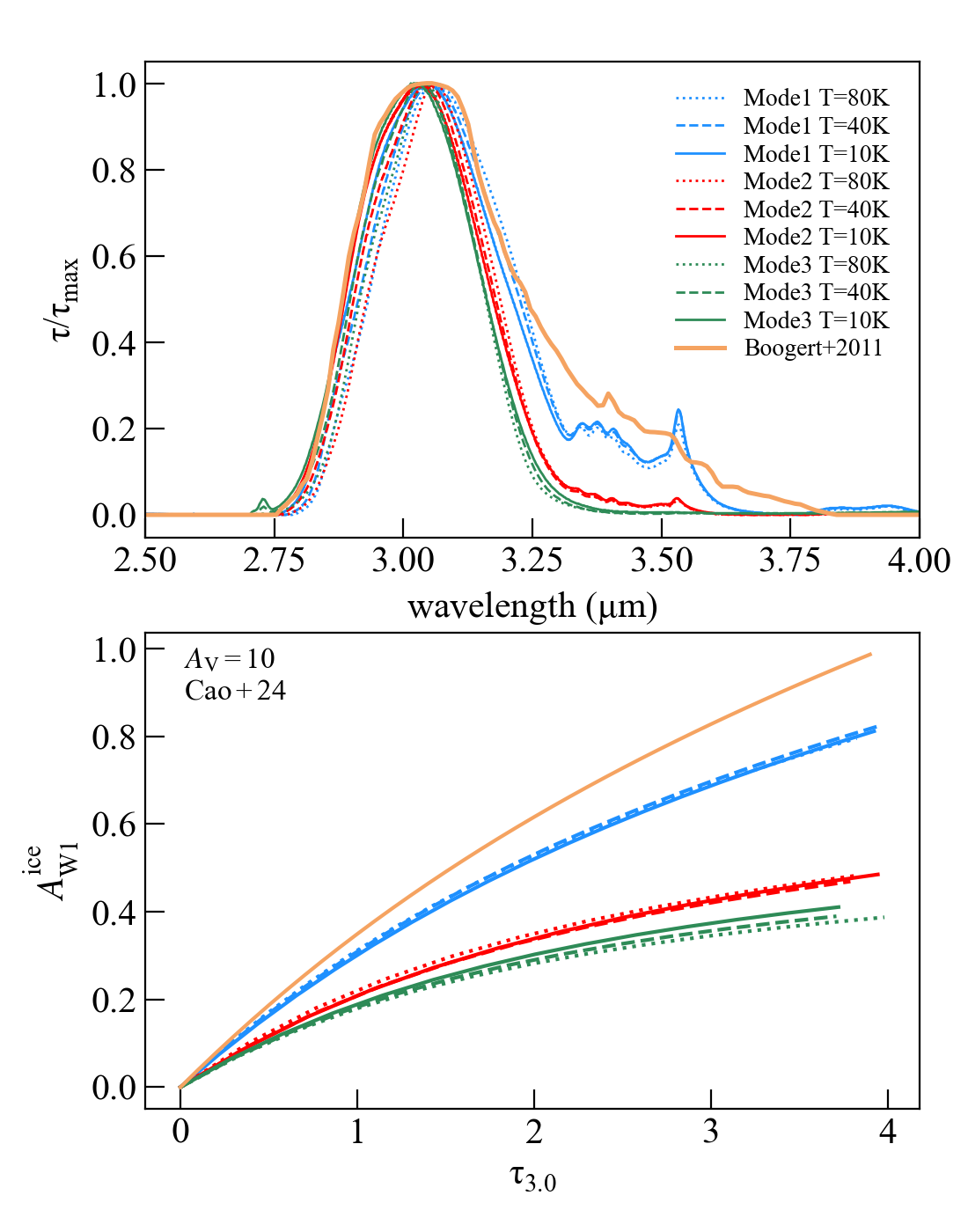}
    \caption{The water ice absorption profiles and the corresponding relationships between $\tau_{3.0}$ and $\Aadd$ derived from each profile. Mode 1, 2, and 3 represent $\rm{H_{2}O:CH_{3}OH:CO:NH_{3}=100:50:1:1}$, $\rm{H_{2}O:CH_{3}OH:CO:NH_{3}=100:10:1:1}$, and the pure water ice profile, respectively. The yellow profile is from the observations of \citet{Boogert11_Spec_DenseCore}.}
    \label{fig:IceProfile}
\end{figure}

We fix the extinction law with Cao+24 and $A_{\rm{V}}=10$ and calculate the synthetic $\Wi$ band water ice absorption under different ice absorption profiles. In Figure \ref{fig:IceProfile}, the upper panel shows the water ice absorption profiles under different conditions taken from the Optical Constants database (OCdb) library (\url{https://ocdb.smce.nasa.gov/}), which provides optical constants of organic refractory materials and ices relevant to planetary and astrophysical environments. Modes 1, 2, and 3 represent $\rm{H_{2}O:CH_{3}OH:CO:NH_{3}=100:50:1:1}$, $\rm{H_{2}O:CH_{3}OH:CO:NH_{3}=100:10:1:1}$, and the pure water ice model, respectively \citep{OCDB_Hudgins93}. Different line styles represent the different dust temperatures. In addition, an observed water ice profile which exhibits the red wing is included for comparison \citep{Boogert11_Spec_DenseCore}. The lower panel shows the synthetic $\Wi$ band absorption profile at different $\tau_{3.0}$ levels. 

Compared with the other factors that will be discussed below, the absorption profile has the greatest impact on the synthetic $\Wi$ band water ice absorption results. This is because the wing adds extra absorption without changing the $\tau_{3.0}$, which is used to measure the amount of water ice. The response of $\Wi$ band reflects that $\Wi$ is much more sensitive in the wing region than at 3 $\mu$m. Therefore, even with the same $\tau_{3.0}$, a difference in absorption profile of wings can produce a large difference in the synthetic $\Wi$ band water ice absorption. 

\subsection{Extinction Law and Extinctions}
Water ice absorption is usually accompanied by dust extinction. When calculating the synthetic $\Wi$ band water ice absorption, the extinction component is mainly affected by two parameters, the shape of the extinction curve and the extinction level. In this subsection, we use the absorption profile from \citet{Boogert11_Spec_DenseCore} and then vary both the extinction law and the extinction level to study the influence on the result.

\begin{figure}
    \centering
    \includegraphics[width=\hsize]{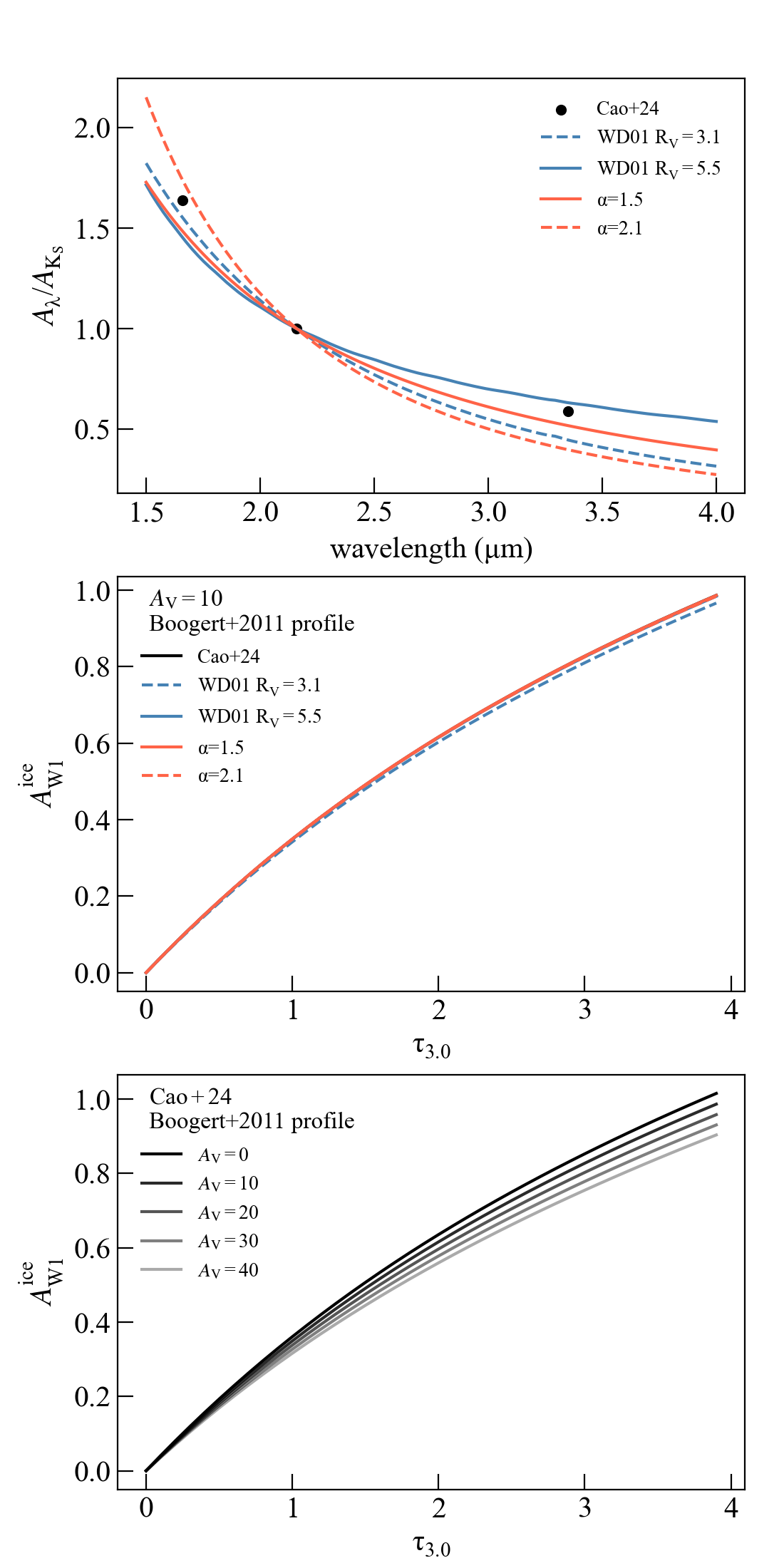}
    \caption{Extinction laws (upper panel) and the corresponding relationships between $\tau_{3.0}$ and $\Aadd$ (middle and lower panels) derived from different extinction laws and different levels of extinction.}
    \label{fig:Ext_and_Law}
\end{figure}

In Figure \ref{fig:Ext_and_Law}, the upper panel shows few IR extinction laws from different models, including the model from \citet{WD01}, the classic near-IR power law extinction law and Cao+24. The middle panel illustrates the influence from the extinction law alone. The curves in the figure show minor differences, indicating that the extinction law has little impact on the resulting $\Aadd$. However, it should be noted that, in Section \ref{sec:Influence}, we pointed out that the extinction law plays a major role when deriving $\Aadd$. These two sections are not in conflict. This is because the extinction law itself has little influence on the resulting $\Aadd$ when we fix other parameters. However, both the extinction measurement and the determination of intrinsic colors strongly depend on the choice of extinction law, which hugely influences the calculation of $\Aadd$.

The lower panel of Figure \ref{fig:Ext_and_Law} shows the influence from extinction. Even with the same $\tau_{3.0}$, the resulting $\Aadd$ changes noticeably with extinction. This variation arises from non-linear photometric effects. Specifically, the convolution between the stellar spectral energy distribution (SED) and the band response is highly complex. The bandwidth of the filters and the shape of the stellar SED can both affect the result, so that the extinction that is monochromatic is non-linear to the extinction expressed with a band in some cases, details can be found in \citet{elephant_extinction}. This non-linear relation has been noticed and applied in papers \citep{Wang19_Law, G_extinction_Danielski18}. Although these non-linear effects are less evident in the IR bands, they can still become significant when the extinction is large. As shown in Figure A1 in \citet{NIR_law_Apellniz20}, both stellar type and extinction level affect the results. In our case, since most stars are red clumps, the SED effect is relatively minor. However, $\tau_{3.0}$ is a parameter to measure the monochromatic amount of extinction, while $\Aadd$ is the band-integrated extinction. The non-linear relationship appears here and will vary with the extinction level as shown in the lower panel in Figure \ref{fig:Ext_and_Law}. We calculate the influence level from extinction on the resulting $\Aadd$. As shown in Figure \ref{fig:Ext_and_Law}, a 1 mag increase in $A_{\K}$ (or an increase of 9.1-12.1 mag in $A_{\rm V}$) leads to an approximately 3\% decrease in $\Aadd$.

Fortunately, since the extinction level correlates with the amount of ice and the resulting $\Aadd$ in this paper is relatively small, the uncertainties introduced by extinction are not evident. Overall, the influence from extinction and extinction law to the synthetic $\Wi$ band water ice absorption are relatively minor compared with the influence on the absorption profile.

\section{Photometric Method Compared with Spectrum Result}\label{sec:Verification}
\begin{figure*}[h!]
    \centering
    \includegraphics[width=\hsize]{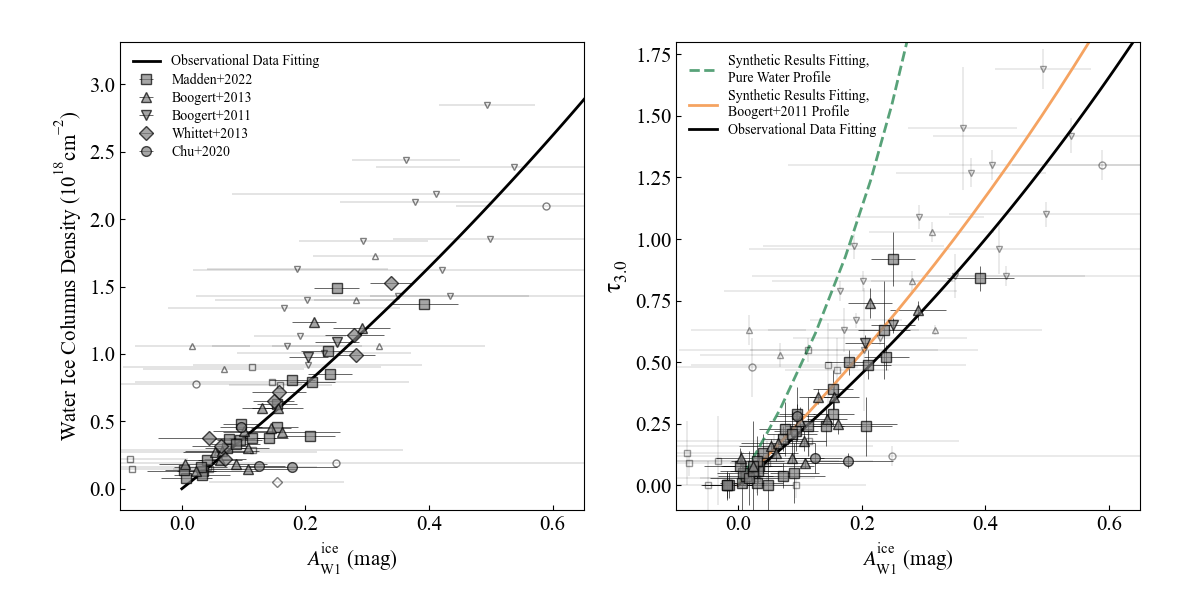}
    \caption{The comparison between the water ice column density (left panel) and optical depth $\tau_{3.0}$ (right panel) from the literature and the $\Aadd$ obtained in this work. Different symbols represent results from different studies, and the bars indicate the uncertainties. A total of 116 stars were cross-matched and are shown as open symbols, among which 70 sources that passed the quality selection in Section \ref{subsec:criterion} ($A_{\K} > 0.3$ was not applied in order to retain more sources) are shown as filled symbols. The green and yellow lines represent the relations predicted from model derived synthetic $\Wi$ band water ice absorption with different absorption profile, profile with red wings (from Boogert+2011) and profile from pure water (no wings). The empirical results from fitting the spectroscopic observation are shown as the black line.}
    \label{fig:SpecCompare}
\end{figure*}

We compare photometry results with spectroscopic observation results, and try to predict $\tau_{3.0}$ from $\Aadd$. Many studies have been done to explore the water ice absorption in the past several years. \citet{Chu20_Spec_COM} chooses 5 small dense molecular clouds with different star forming environments, from low-mass star formation region to two starless collapsing cores. \citet{Whittet13_Spec_L183} uses 9 stars in pre-stellar dark cloud L183, \citet{Boogert11_Spec_DenseCore, Boogert13_Spec_Lupus} trace the quiescent medium of isolated dense cores and compare the ice formation and grain growth in Lupus molecular cloud, \citet{Madden22_Spec_PerSer} uses 49 field stars behind the Perseus and Serpens Molecular Clouds to find the relations between extinction and water ice and silicate absorption. In Figure \ref{fig:SpecCompare}, the results from the papers above are all collected and cross-matched with the photometry results to compare and evaluate consistency of the two methods. Different markers representf results from different papers, and all 70 out of the 116 stars that passed the criterion in Section \ref{subsec:criterion} are highlighted using solid markers. The criteria are about extinction quality selection from Equation \ref{eq:AkSelection} and $\overline{D}_{\mathrm{neighbors}}$ selection from Equation \ref{eq:IntrinsicSelection}. In order to retain more stars, the $A_{\K}>0.3$ criterion is removed, otherwise, only 59 stars would remain. In Figure \ref{fig:SpecCompare}, the relationship between $\Aadd$ derived in this work and the water ice abundance measured from spectroscopy is shown. The left panel compares $\Aadd$ with the water ice column density, and the right panel compares it with the optical depth. The synthetic $\Wi$ band absorption described in Section \ref{sec:SynPhot} is also included. The green and orange curves represent the results obtained using two different absorption profiles, the \citet{Boogert11_Spec_DenseCore} profile (orange) and the Mode3 40K profile (pure water at 40 K, green) described in Section \ref{sec:SynPhot}. 

As discussed in Section \ref{subsec:profile}, the absorption profile plays the dominant role in determining the relationship between $\tau_{3.0}$ and $\Aadd$, particularly the shape in red wings. In the literature, only a small number of stars show no wing \citep{Boogert11_Spec_DenseCore, Boogert13_Spec_Lupus, Madden22_Spec_PerSer, RedWings_Thi06}, whereas most background stars, more or less, exhibit red wing structures. Therefore, we adopt the profile from \citet{Boogert11_Spec_DenseCore}, which includes the red wing, and set the extinction to $A_{\rm{V}}=10$ with Cao+24 extinction law to present a typical extinction in this work. The relationship between $\tau_{3.0}$ and $\Aadd$ with \citet{Boogert11_Spec_DenseCore} profile is plotted in orange color in Figure \ref{fig:SpecCompare}. The green curve which represents the result by using Mode3 40K profile (pure water at 40 K) described in Section \ref{sec:SynPhot} are also plotted as an upper limit for the $\tau_{3.0}$ estimation. An exponential function is applied to fit the relationship between $\Aadd$ and $\tau_{3.0}$ for these two synthetic results. The fitting results are as follows,
\begin{equation}
\begin{aligned}
\tau_{3.0}=2.5594(\rm{e}^{0.9383\times\Aadd}-1)
\end{aligned}
\label{eq:SynFitB11}
\end{equation}
This equation is the result with Boogert+2011 profile (with red wings) and model derived synthetic absorption, as shown in right panel of Figure \ref{fig:SpecCompare} with orange solid line.
\begin{equation}
\begin{aligned}
\tau_{3.0}=0.4770(\rm{e}^{5.6297\times\Aadd}-1)
\end{aligned}
\label{eq:SynFitPureWater}
\end{equation}
This equation is the result with pure water profile (no red wings) and model derived synthetic absorption, as shown in right panel of Figure \ref{fig:SpecCompare} with green dash line.

We also applied the fitting to the observational spectroscopic data, using measurements from 70 stars. The fitting represents an empirical relation derived purely from observational data. The relationship follows:
\begin{equation}
\begin{aligned}
\tau_{3.0}=2.2399(\rm{e}^{0.9231\times\Aadd}-1)\\
N/(10^{18}\rm{cm^{-2}})=6.1812(\rm{e}^{0.5894\times\Aadd}-1)
\end{aligned}
\label{eq:SynFitObservation}
\end{equation}
The equation is from empirical fitting with spectroscopic observation results, as shown in Figure \ref{fig:SpecCompare} with black solid line.

Whether in the literature or in model predictions, the water ice column density $N$ and the optical depth $\tau_{3.0}$ remain highly correlated, essentially following a linear relationship, which can be expressed as:
\begin{equation}
N/(10^{18}\rm{cm^{-2}})= 1.635\times\tau_{3.0}
\label{eq:tau2N}
\end{equation}

\section{Ice Distribution Map}\label{sec:Result}
\subsection{From $\Aadd$ to Ice Distribution Maps}
To construct a map based on the $\Aadd$ results above, we use HEALPix to generate a regular, equal-area pixelization of the sky. We set $N_{side}=2^{9}$, which produces approximately 3.1 million pixels (a pixel size of about $7^\prime$). Each pixel maps to a small angular area on the sky, and the pixel centers are used as reference positions for smoothing and constructing the water ice distribution map. For each pixel center, a smooth scale parameter of $\gamma=0.1^\circ$ is applied which is close to the resolution. Tests with different $\gamma$ values show no significant impact on the results. Besides, only stars within an angular radius of $3\gamma=0.3^\circ$ are included and a minimum of 5 stars within the radius is also required. Errors together with the angular distance to the pixel center are adopted as weights for each star:
\begin{equation}
\begin{split}\label{eq:SmoothMap}
w_{i} = \frac{1}{\sigma_{A_{\Wi,i}^{\rm{ice}}}^{2}} \times \exp (-\frac{D_{i}^{2}}{2\gamma^{2}})\\
A_{\Wi}^{\rm{ice, smoothed}} = \frac{\sum w_i\times A_{\Wi,i}^{\rm{ice}}}{\sum w_i}
\end{split}
\end{equation}
where $\sigma_{A_{\Wi,i}^{\rm{ice}}}$ is the uncertainty of $\Aadd$ for each tracer star $i$, $D_{i}$ is the angular separation between tracer star $i$ and the pixel center, and $\gamma$ is a smoothing scale parameter. The smoothed value at the pixel center $A_{W1}^{\rm{ice,smoothed}}$ is thus obtained through a weighted average of nearby sources.

\subsection{General Ice Distribution Map}\label{GenDis}
\begin{sidewaysfigure*}
  \centering
  \includegraphics[width=\textwidth]{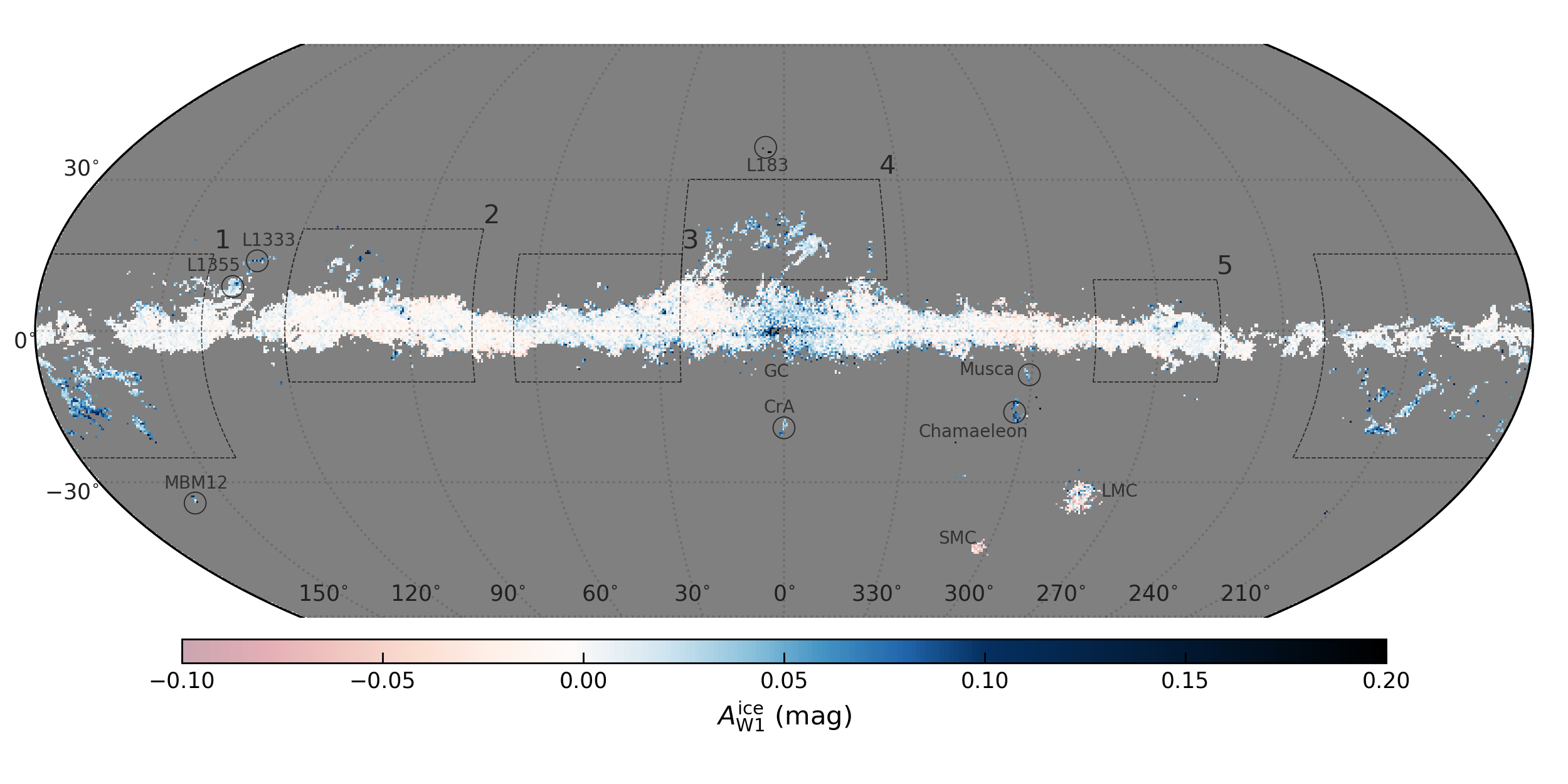}
  \caption{Distribution of water ice. The color map represents the smoothed $\Aadd$ values, where blue indicates the existence of water ice absorption. Several well-known regions are enclosed in boxes and labeled as regions 1 to 5, they are (1)Orion-Taurus-Perseus (OTP) region, (2)Cygnus Complex, (3)Serpens region, (4)Ophiuchus region and (5)Vela Complex. Details about these regions are described in Section \ref{subsec:DetailDis}. In addition, some other famous clouds with water ice signals are labeled.}
  \label{fig:Distribution}
\end{sidewaysfigure*}

The water ice distribution is shown in Figure \ref{fig:Distribution}. Due to the selection described in Section \ref{sec:Calculation}, it is important to note that the map primarily traces water ice absorption in regions where $A_{\K}>\max(0.3, 3\sigma_{A_{\K}})$. Also, the 2MASS and WISE data used in this study have an extinction detection limit of about $A_{\K}<3$ mag ($A_{\K}/A_{V}\approx0.08-0.11$, \citet{Wang19_Law, WD01}), which limits both the depth and spatial coverage. The resolution of the map is about $7^\prime$. The extinction limit and the photometry depth from 2MASS survey restrict the exploration of distant clouds. Most of the patterns in the map are from the nearby cloud. Deeper photometric data, such as from Spitzer, would allow a more detailed analysis of distant molecular clouds.

The $\Aadd$ values of tracer stars follow an approximately normal distribution with a mean of $-0.003$ and a standard deviation of $0.057$, while the smoothed $A_{\rm W1}^{\rm ice, smoothed}$ in the ice map have a mean of $0.005$ and a standard deviation of $0.027$. We therefore recommend $3\sigma_{\Aadd}\sim[0.16, 0.18]$ is used as the threshold for tracer stars with ice absorption, while $3\sigma_{A_{\rm W1}^{\rm ice, smoothed}}\sim[0.08, 0.09]$ is recommended as the threshold for the pixels in the smoothed map to show ice absorption.

The gray areas in Figure \ref{fig:Distribution} indicate regions with extinction below the threshold $A_{\K}<0.3$. The color map represents the distribution of the smoothed $\Aadd$. The deep blue pixels represent robust water ice absorption and correlate well with molecular cloud structures. Most of the faint red/blue pixels can be attributed to calculation or observational uncertainties. However, the light blue pixels may also arise for other reasons. A weak ice absorption can also result in light blue pixels. Moreover, the surrounding filament or dark core structures typically have small angular sizes and their absorption strength is likely to be smoothed out when deriving the ice distribution map, such as the sub figure A in Figure \ref{fig:TPO}. In addition, if these structures lie in the Galactic plane, the surrounding noise may further weaken their water ice absorption signals into pale blue. It is difficult to extract such smoothed or weak absorption features from the noise at the current precision. New methods are needed to improve the detection sensitivity.

We carefully inspected the entire map and identified the regions showing ice absorption and listed them in Appendix \ref{App:Table} based on the following selection: (1) the presence of a sufficient number of stars exhibiting strong water ice absorption; (2) clustering of these stars within dense regions with extinctions; and (3) a corresponding signal in the smoothed ice map. The list in Appendix \ref{App:Table} is not intended to be a complete catalog of nearby icy clouds, since some regions may be missed due to line of sight overlap or small angular sizes. However, all regions in the list are verified with ice absorptions.

\subsection{Detailed Ice Distribution Map}\label{subsec:DetailDis}

We will discuss the ice absorption in some regions with clear ice absorption signal, providing a more localized view of ice distribution. The regions that will be discussed are shown in Figure \ref{fig:TPO}, \ref{fig:Cyg} and \ref{fig:DetailOthers}. In some specific regions, we plot all the tracer stars, with three different markers used to indicate the three different levels of $\Aadd$, $\Aadd<0.05$, $0.05<\Aadd<0.15$ and $\Aadd>0.15$, over an image background from AKARI Far-IR Surveyor or DSS \citep{AKARI,ALADIN} to show the distribution of ISM. The smallest dots correspond to $\Aadd < 0.05$ and are considered stars without ice absorption (approximately $\sigma_{\Aadd}$). The blue triangles represent $0.05 < \Aadd < 0.15$, indicating stars that are likely to exhibit ice absorption. The green stars mark objects with $\Aadd > 0.15$, which indicates clear ice absorption features. Given the map resolution of approximately 7' and the selection criteria described above, we estimate that icy clouds with angular sizes larger than about 15' are likely to be detected, provided a sufficient number of tracer stars are present in the region. Smaller features, particularly filament substructures and dense cores, may fall below this limit and be smoothed out. We note, however, that this estimate depends strongly on the local tracer star density, and regions with sparse sampling may require even larger angular extents for a robust detection. We select and list some of the clouds that exhibit significant water ice absorption.

\subsubsection{Region 1: Taurus-Orion-Perseus}

\begin{figure*}[h!]
  \centering
  \includegraphics[width=\hsize]{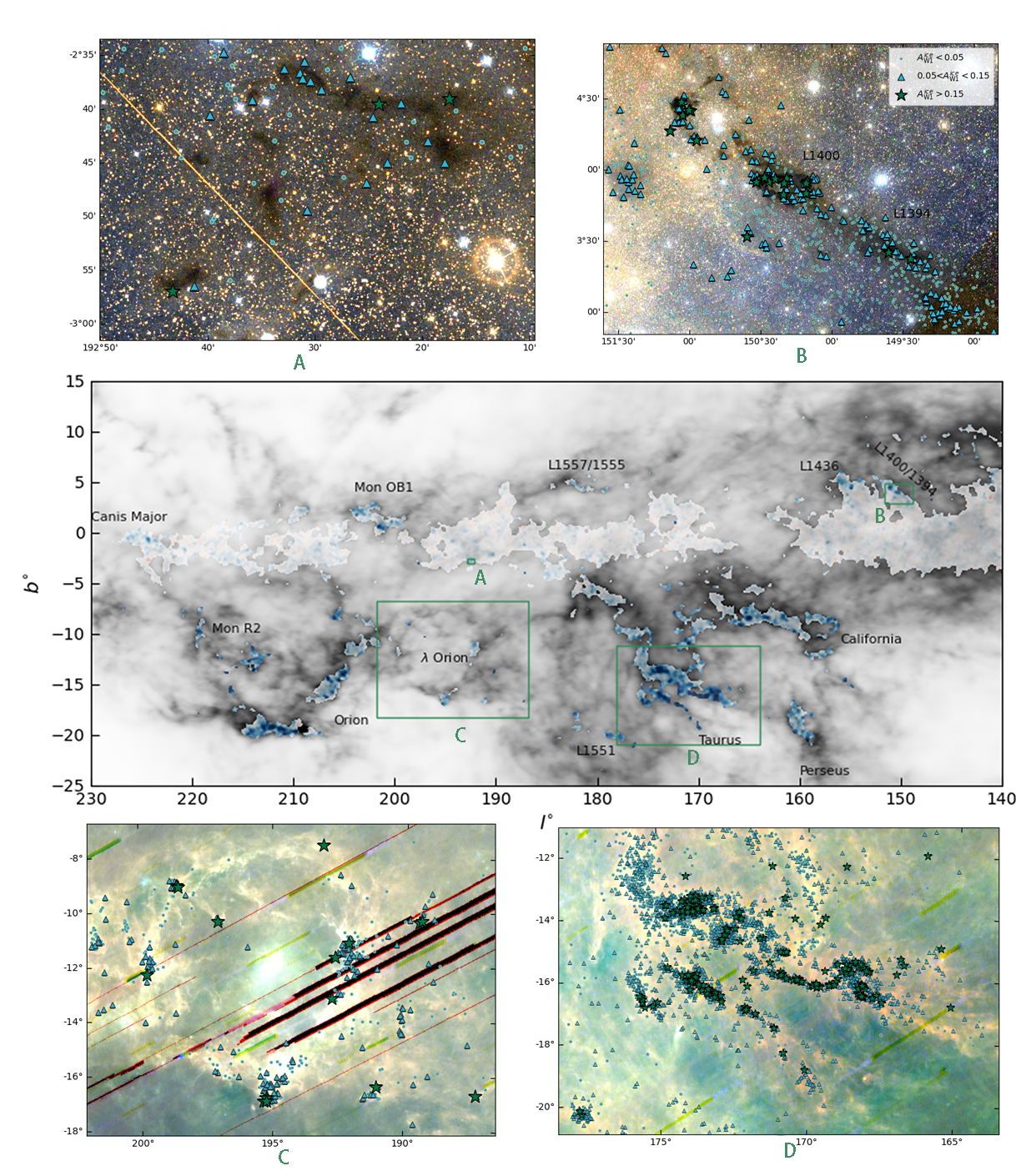}
  \caption{The detailed map of Taurus-Orion-Perseus region. The background of the middle panel is the dust map from \citet{dustmap_GE24}, indicating the extinction level (note that this map only covers the distance around 1.25 kpc). Blue colors represent a higher water ice absorption. Locations with clear water ice detections are labeled. The four panels surrounding the middle panel are magnified views of the four regions, A, B, C and D in the mid panel, and each plot shows all the tracer stars, with three different markers used to indicate the three $\Aadd$ levels, $\Aadd<0.05$, $0.05<\Aadd<0.15$ and $\Aadd>0.15$. The backgrounds for top two panels are from DSS2, while the backgrounds for the bottom two panels are from AKARI. Sub figures A, B and C show the stars with ice absorption in filament structure or core region with small angular sizes which are easily smoothed out in ice distribution map, while D shows the stars with ice absorptions in the large scale Taurus molecular cloud. The color bar of the mid panel is shared with Figure \ref{fig:Distribution}.}
  \label{fig:TPO}
\end{figure*}

The mid panel in Figure \ref{fig:TPO} represents a region containing several well studied molecular cloud complex, including Taurus, Perseus, Orion and California. All of them are found to be rich in water ice. The sub figure D in Figure \ref{fig:TPO} shows the spatial distribution of stars exhibiting water ice absorption in Taurus, with an AKARI image as the background. Stars with clear water ice absorption are strongly correlated with the dust distribution and tend to cluster together. It is observed that mid-level ice absorption is more prevalent on the left side of the Taurus molecular cloud compared with that on the right side, but the reason is unknown. For other cloud complex, such as Canis Major and Mon OB1, which are more distant and whose angular sizes are correspondingly smaller, only a limited number of blue pixels mark the evidence of water ice absorption. 

The surrounding filament or dark core structure is somewhat complex to interpret. An example from L1400/1394 is discussed here for filament structure. In sub figure B, while L1400/1394 clearly shows the existence of water ice in dense regions in the scatter map with tracer stars, the feature becomes less visible in the ice map because of smoothing and its small area. A similar case is in Canis Major. An even more extreme example is in region A, where the noisy environment, small angular size, and weak absorption make the possible ice signal entirely invisible in the ice distribution map. Similar to filament structures, cores have even smaller angular sizes and fewer star tracers, making them more likely to be smoothed out in the ice distribution map.

\subsubsection{Region 2: Cygnus Complex}

\begin{figure}
  \centering
  \includegraphics[width=\hsize]{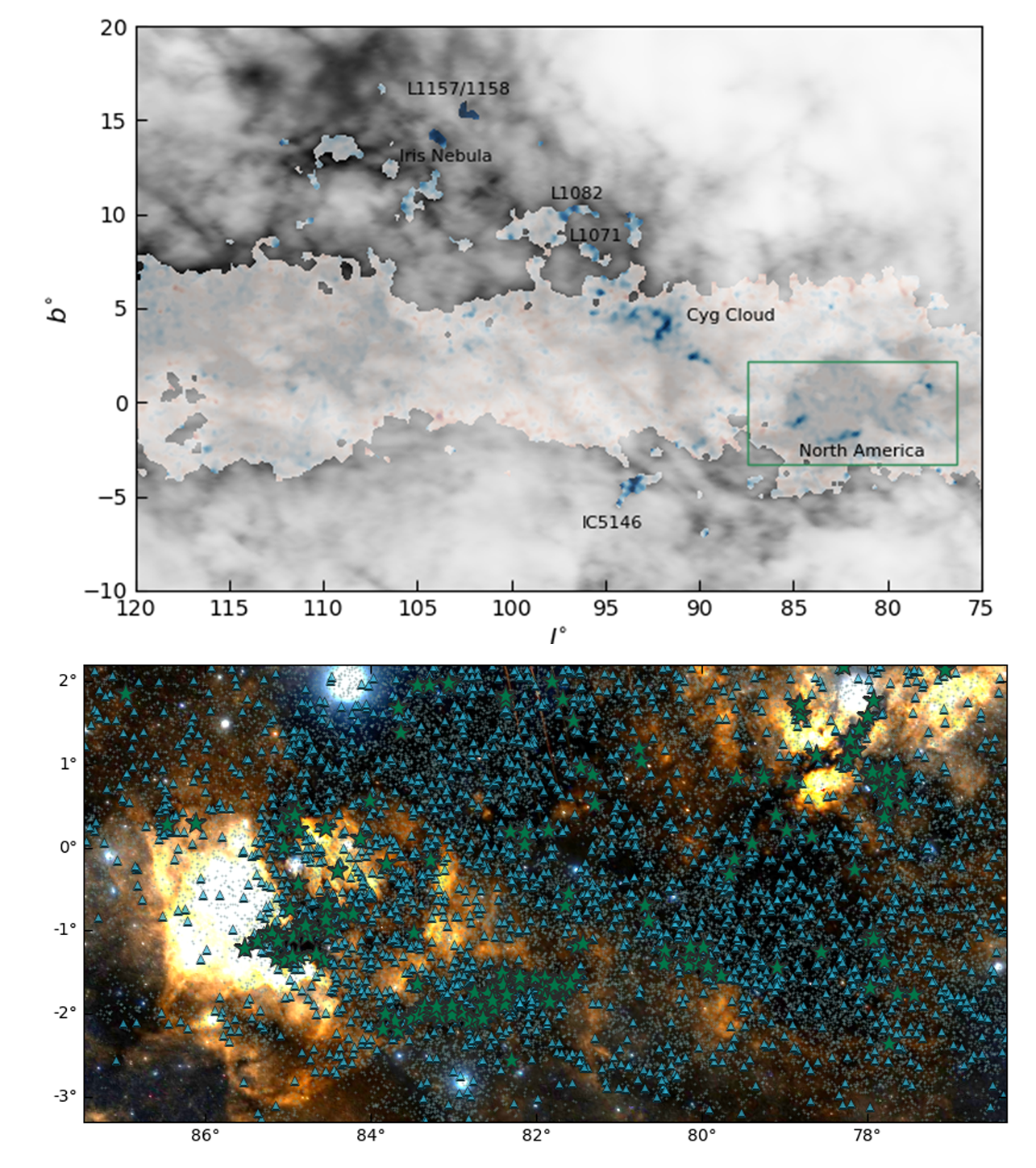}
  \caption{Same as Figure \ref{fig:TPO}, but for the Cygnus Complex region. The upper panel is the dust map with ice map, the lower panel is for the North America Nebula which has a complex environment in the Galactic plane making it difficult to distinguish the ice signal from the noise. The color bar of the upper panel is shared with Figure \ref{fig:Distribution}.}
  \label{fig:Cyg}
\end{figure}

In Figure \ref{fig:Cyg}, the upper panel shows the water ice distribution of Cygnus Complex. From the 3D dustmap \citep{dustmap_GE24}, it is obvious that multiple clouds overlap along the line of sight in this region. The panel indicates the presence of water ice, however, it is difficult to attribute the water ice to any specific cloud. 

Clear water ice absorption is found in several dark clouds away from the Galactic plane (such as L1157, L1158, L1172, L1174, L1082, and L1071, as marked in the figure). Within the Galactic plane, a strong and extended water ice absorption feature is detected at $l=(90^{\circ},95^{\circ})$, $b=(3^{\circ},5^{\circ})$. We refer to this structure as the `Cyg Cloud'. IC 5146 is a nearby dark cloud complex in Cygnus (about 200 pc). Some studies focus on the chemistry inside the cloud and water ice is clearly observed in the spectrum \citep{IC5146_Chiar11}.

It is also obvious that the North America Nebula exhibits a clear water ice absorption feature. In addition, the right side of the North America Nebula ($l=(75^{\circ},80^{\circ})$) 
seems to have a weak absorption signal in the water ice distribution map. However, it is difficult to determine whether these light blue patterns truly originate from water ice because of the complex environment and the noise level. The lower panel in Figure \ref{fig:Cyg} shows the complex conditions in North America Nebula. On the right side of the nebula, the clouds overlap, and stars with different ice absorption levels are mixed making it comparatively chaotic. Therefore we simply note the existence of water ice in North America Nebula.

\subsubsection{Region 3: Serpens Complex}
\begin{figure}
  \centering
  \includegraphics[width=\hsize]{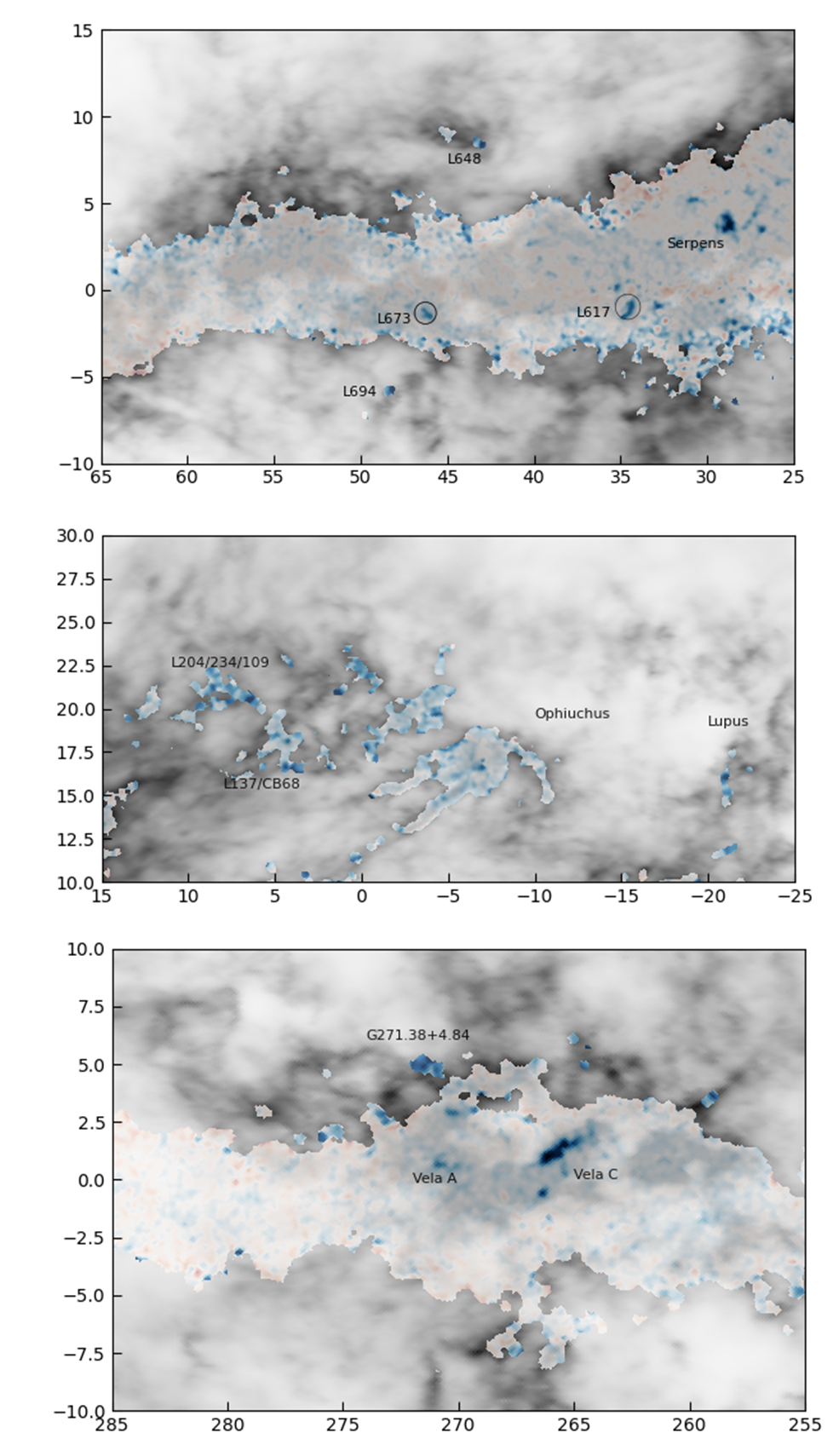}
  \caption{Same as Figure \ref{fig:TPO}, showing the detailed maps of Serpens region (upper panel), Ophiuchus region (middle panel) and Vela Complex region (lower panel). The color bar of the 3 panels is shared with Figure \ref{fig:Distribution}.}
  \label{fig:DetailOthers}
\end{figure}

Since Serpens region is located closer to the Galactic Center, the ISM becomes more abundant, the situation becomes more complex, and the fluctuations and noises of the background are more obvious. In this area, in order to verify the existence of water ice correctly, we not only examine the ice distribution map and the tracer star distribution map, but also check whether the stars with water ice absorption show any clustering and whether they associate with molecular clouds in DSS2 and AKARI image. Based on these criteria, we identify several clearly absorbing regions, which are marked in Figure \ref{fig:DetailOthers}.

Serpens molecular cloud is one of the most prominent regions with a large area of widespread water ice absorption across the sky. Spectroscopic observations also verify the presence of water ice in Serpens \citep{Madden22_Spec_PerSer}. The main cloud is shown as the deep blue area on the right side of the upper panel in Figure \ref{fig:DetailOthers}. Also, several connected filament structures around it show water ice absorption, e.g., the filament on the right-hand side of the deep blue area. Apart from Serpens, in this field, there are cores located far from the Galactic plane, which are free from the complexity of the Galactic plane. They can be identified even with small angular sizes, as in L648 and L694. \citet{Chu20_Spec_COM} confirms the presence of clear water ice absorption in L694 with spectrum. However, for targets on the Galactic plane, due to contamination, objects must be larger to be detected, such as L673 and L617, and \citet{Boogert11_Spec_DenseCore} includes water ice observations toward L673.

\subsubsection{Region 4 and 5: Ophiuchus and Vela Complex}
The middle panel in Figure \ref{fig:DetailOthers} shows that water ice is abundant in Ophiuchus and Lupus. This is similar to Taurus and Orion molecular clouds which are close to The Earth and far from the Galactic plane. In most situations, Ophiuchus can be considered as a single cloud along the line of sight. However, compared with other single molecular clouds such as Taurus, its water ice absorption is relatively weak. We attempted to explain this by different radiation fields and temperatures associated with star forming region, but Orion also exhibits strong water ice absorption, which challenges this interpretation. Since the existence of water ice requires specific physical conditions, we plan to investigate next whether the 3D distribution of molecular clouds is responsible for this discrepancy.

The lower panel in Figure \ref{fig:DetailOthers} shows the water ice absorption in Vela Complex. Vela is also one of the most prominent regions with water ice absorption in the Milky Way. The situation near Vela is comparatively simple, since from the dust distribution in the Milky Way, Vela is one of the few nearby molecular clouds around $l\approx270^{\circ}$.

\subsubsection{Other Clouds, Magellanic Cloud and Galactic Center}
Some other nearby clouds are also presented in Figure \ref{fig:Distribution}, such as Corona Australis, Chamaeleon and Musca. In addition, some filaments or dark core structures, L1355, L1333, L183 \citep{Whittet13_Spec_L183} and Perseus Arm 1 \citep{Cao23_structure} (MBM 12), also have water ice absorption. 

In addition, the ice tracers also include a number of stars in the Large Magellanic Cloud (LMC, $(l,b)\sim(280.5^{\circ},-32.9^{\circ})$) and the Small Magellanic Cloud (SMC, $(l,b)\sim(302.8^{\circ},-44.3^{\circ})$), but their water ice absorption shows no clear spatial pattern. Considering the distance to the LMC and the extinction, only very bright red giants can be detected by 2MASS and WISE. However, the intrinsic colors of red giants are difficult to determine accurately, leading to large uncertainties or even errors in the derived absorption. Therefore, the LMC and the SMC are not discussed in this paper. Nonetheless, with higher resolution and deeper photometric data, exploring water ice absorption distribution outside the Milky Way would be an interesting project.

The Galactic Center region is complex, and although strong water ice absorption is detected there, the limited resolution and the complicated local environments and cloud structures prevent a detailed analysis. We therefore only note the presence of strong absorption in this region.

\section{Discussion}\label{sec:Discussion}
The photometric method opens a new window for exploring interstellar ice, as it requires less data: photometric measurements in the W1 band, extinctions and intrinsic colors. (1) Small scale structures of some regions with water ice absorption can be discovered, which are generally inaccessible through previous observations and can only be revealed by map surveys. (2) Map based studies. The ice map can naturally be compared with other maps, such as dust temperature maps, radiation maps, or any tracers expected or suspected to be correlated. Also, integrating the 2D water ice map to derive the `water reservoir' of some molecular clouds provides observational constraints relevant to elemental budgets, such as the long-standing `O crisis', namely that the total number of O atoms depleted from the gas phase far exceeds that tied up in solids \citep{Wang15_Ocrisis}. (3) Complementarity with spectroscopy for further possibilities. For instance, Spitzer IRS \citep{SpitzerIRS} covers the range from 5.3 to 38 $\mu$m, including silicate and other features but not the ice feature, the combination enables the identification of the potential connections between molecular ices.

Beyond the high-coverage but relatively low-sensitivity 2MASS data in this work, deeper and higher-resolution surveys from telescopes such as Visible and Infrared Survey Telescope for Astronomy \citep{VISIONS, VVV} (VISTA), United Kingdom Infra-Red Telescope \citep{UKIRT} (UKIRT), or Spitzer \citep{Spitzer} can be utilized. This is particularly valuable because $\Wi$ is the key photometric band, and the J band suffers extinction four to six times stronger than W1 band \citep{Wang19_Law, WD01}, observations are often limited by the shorter wavelength bands, in this case, J band limits the observation depth even when the W1 band has not yet reached its limiting magnitude. Therefore, deeper and higher resolution photometry would better exploit the W1 data and enable a more thorough exploration of these extremely extincted regions. In the JWST era, the method could be applied not only to water ice but also to other molecular ices \citep{PhoIce_Ginsburg25}. \citet{Ginsburg23_CO} has demonstrated that other molecular ice absorptions, such as CO, can also be studied with the photometric method. With the various filters available on JWST, images alone are sufficient to construct spatial distribution maps of different molecules, allowing direct study of spatial patterns.

\section{Summary}\label{sec:Summary}
\begin{itemize}
    \item We map the distribution of water ice, with extinction and intrinsic color derived from PNICER method and stellar locus method. The ice absorption $\Aadd$ in W1/WISE band is calculated and high quality tracer stars are selected to construct the first water ice distribution map across the Milky Way. 
    \item The extinction law has a significant impact on $\Aadd$. The extinction law affects both the estimation of extinction and intrinsic colors, and will introduce an offset on the resulting $\Aadd$.
    \item By combining different extinction conditions and water ice absorption profiles, we use synthetic photometry to investigate the factors that influence the theoretical $\Aadd$ most. The results show that the absorption profile (e.g., a red wing or other molecular absorption) has the largest impact on $\Aadd$. In addition, the nonlinear extinction effect also influences $\Aadd$ but only to a limited extent.
    \item We compare $\Aadd$ from photometric method with spectroscopic observation results, and find a strong correlation between the two. A theoretical estimate and an empirical relation of the relationship between $\Aadd$ and $\tau_{3.0}$ or water ice column density is provided. The relation between the water ice column density $N$ and $\tau_{3.0}$ is also presented.
    \item An ice distribution map is established with a resolution of around $7^\prime$. Regions with clear water ice absorption are marked in both the general map in Figure \ref{fig:Distribution} and are discussed in detail below. The detection area is limited in extinction, with a minium and maximum depth in $A_{\K}$ being 0.3 and 3, $A_{V}\sim[3,30]$ ($A_{\K}/A_{V}\approx0.08-0.11$).
\end{itemize}

\section*{Data Availability}
The dataset of 618k tracer stars, including $\Aadd$, photometry, and coordinates, as well as the water ice map HEALPix file, are provided online. They can be accessed at XXXXX.

\begin{acknowledgements}
We are grateful to Prof. Jo\~ao Alves, Dr. Martin Piecka, Mr. Hongrui Gu, Dr. Jun Li and Dr. Mingxu Sun for their helpful assistance and suggestion. This work is supported by the NSFC project 12133002 and China Scholarship Council (Grant No.202406040156). This work has made use of the data from 2MASS and WISE surveys.
\end{acknowledgements}

\bibliographystyle{aa}
\bibliography{reference}

\begin{appendix}
\section{Reasons of Using Giants as Control Samples and the Set of Extinction Threshold}\label{App:tthSet}
In \citetalias{stefan17_PNICER}, as pointed out in Section 3.3, a degeneracy can occur when a reddened star is moved along the extinction vector using the PNICER method. Some stars may intersect multiple components in the intrinsic CCD (e.g., the galaxy locus, the M-sequence dwarf branch, and early-type stars), leading to multiple peaks in the extinction probability density distribution. These multiple peaks make it difficult to determine the true extinction and intrinsic color of the star. In this work, these multiple peaks arise from the dwarfs and giants components in the intrinsic CCD. 

As shown in the left panel of Figure \ref{fig:IC_CCD}, the comparison between the intrinsic color distributions of dwarfs and giants clearly shows that the dwarf intrinsic color distribution exhibits a noticeable `spur' feature. This feature arises because dwarfs have higher surface gravity than giants, leading to more frequent collisions between atoms and a higher probability of molecule formation. The enhanced molecular formation in dwarfs strengthens molecular absorption in the infrared, making their intrinsic colors systematically different from those of giants \citep{IRcolorStars_Allard95, IRcolorStars_Bessell88, RSG_Ren21}. In this work, this leads to the confusion between low-extinction dwarfs and high-extinction giants, resulting in incorrect extinction estimates and, consequently, errors in the derived intrinsic colors and $\Aadd$. This error in calculating $\Aadd$ can be clearly reflected in the distribution map in the initial results. After plotting the distribution map of stars with significant ice absorption in the initial results, the map shows that a number of stars are uniformly spread across high galactic latitude regions where the extinction is nearly zero. Cross-matching with the APOGEE and Gaia datasets later revealed that most of these uniformly distributed contaminants are foreground dwarfs with relatively large surface gravity (log $g$) and closer distances. 

The solution to this problem is straightforward. Several selection criteria are applied to obtain a control sample and a tracer star dataset, both of which consist primarily of giants. Specifically, (1) we only use `extinction free' giant stars in control sample, with $E_{\mathrm{J},\K}<0.02$, log $g<3.4$ and $\mathrm{ T_{eff}}<5400$K from APOGEE samples. (2) we mitigate the influence from dwarfs by applying an extinction threshold. Since dwarfs are generally much fainter than giants, most of them can not suffer from significant extinction. Therefore, applying an extinction constraint can effectively remove contamination from low-extinction dwarf stars. In this work, after examining the ice distribution maps under different extinction thresholds, we finally adopt $A_{\K} > 0.3$ as a selection criterion to reduce the fraction of dwarfs in the final sample.

We validated our giant star selection method using APOGEE stars with a simple criterion of $\log g = 3.5$ to separate dwarfs and giants. The initial sample contains 229k dwarfs and 358k giants. After applying the same selection procedure above, only 408 dwarfs and 32936 giants remain, increasing the giant purity from 60.99\% to 98.78\% and demonstrating that the method effectively removes dwarf.

\section{Cloud with Significant Water Ice Absorption}\label{App:Table}
The clouds with clear water ice absorption are listed in this section, but the list is not a complete list of the icy clouds in the Milky Way. Some clouds with ice are difficult to identify, either because of low signal strength or the small number of available samples.
\begin{table}[h!]
\caption{Selected Clouds with Water Ice Absorption}
\label{tab:Table}
\centering
\begin{tabular}{lccl}
\hline\hline
name          &$l^{\circ}$ &$b^{\circ}$  & Description \\
\hline
L137/CB68       & 4.4     & 17.8    & e       \\
L183            & 6.067   & 36.757  &         \\
L204/234/109    & 7.5     & 21.0    & e         \\
Serpens         & 28.8    & 3.6     & e,S,L,$\star$    \\
L617            & 34.573  & -0.862  & S        \\
L648            & 43.024  & 8.374   &           \\
L673            & 46.264  & -1.330  & S      \\
L694            & 48.351  & -5.725  &          \\
North America   & 84.7    & -1.0    & e,S,L          \\
Cyg Cloud       & 92.0    & 3.9     & e,S,L,$\star$      \\
IC5146          & 94.382  & -5.520  & S         \\
L1071           & 96.004  & 8.104   &          \\
L1082           & 96.973  & 10.074  &           \\
L1157/1158      & 102.654 & 15.485  &           \\
Iris Nebula     & 104.062 & 14.193  &           \\
L1333           & 128.878 & 13.707  &        \\
L1355           & 133.459 & 8.678   & S,L       \\
L1400/1394      & 149.930 & 3.638   & S         \\
L1436           & 155.720 & 5.034   &          \\
MBM12           & 159.351 & -34.323 &           \\
Perseus         & 160.5   & -18.5   & e,S,L   \\
California      & 161.5   & -8.5    & e,L         \\
Taurus          & 171.5   & -15.5   & e,S,L,$\star$   \\
L1551           & 178.857 & -20.014 &           \\
L1555           & 179.298 & 4.176   &           \\
L1557/L1555     & 181.332 & 4.407   &           \\
$\lambda$ Orion & 195.0   & -12.0   & e         \\
Mon OB1         & 202.08  & 0.984   & S,L         \\
Orion           & 211.0   & -19.5   & e,S,L,$\star$   \\
Mon R2          & 213.701 & -12.596 &         \\
Canis Major     & 224.0   & -2.0    & e,L         \\
Vela C          & 265.3   & 1.5     & e,S,L,$\star$   \\
Vela A          & 270.8   & 0.7     & e,S,L         \\
G271.38+4.84    & 271.380 & 4.840   & S          \\
Chamaeleon      & 300.0   & -15.5   & e,S   \\
Musca           & 300.6   & -0.8    & e         \\
Lupus           & 339.1   & 15.6    & e        \\
Ophiuchus       & 353.2   & 16.5    & e,L        \\
CrA             & 359.8   & -18.2   & e          \\
\hline
\end{tabular}
\tablefoot{The list of the icy cloud and the brief description. \\ 
e: The molecular cloud is too extended to be represented by a single point, or the structure is composed of multiple objects, so only a rough coordinate is given. For $\lambda$ Orion, the center of the ring is given \citep{Cao23_structure}.\\
S: Strong water ice absorption. It is required that at least four to five stars with $\Aadd > 0.25$ be present in the cloud.\\
L: A large area of water ice absorption, based solely on visual inspection, so it is approximate.\\
$\star$:  Prominent water ice absorption\\
}
\end{table}
\end{appendix}

\end{CJK*}
\end{document}